\documentclass{elsart2}
\usepackage{amssymb}

\begin{document}
\begin{frontmatter}

\title{On elliptic solutions of the cubic
complex one-dimensional Ginzburg--Landau equation}

\author{S.~Yu.~Vernov}

\address{Skobeltsyn Institute of Nuclear Physics, Moscow State
University,\\ Vorob'evy Gory,  Moscow, 119992, Russia}

\ead{svernov@theory.sinp.msu.ru}
\date{}

\begin{abstract}
The cubic complex one-dimensional Ginzburg--Landau equation is
considered.  Using the Hone's method, based on the use the
Laurent-series solutions and the residue theorem, we have proved
that this equation has neither elliptic standing wave nor elliptic
travelling wave solutions. This result amplifies Hone's result,
that this equation has no elliptic travelling wave solutions.
\end{abstract}

\begin{keyword}
Standing wave \sep elliptic function \sep residue theorem \sep the
cubic complex one-dimensional Ginzburg--Landau equation \PACS
05.04.-a \sep 02.30.-f \sep 02.70.Wz \sep 47.27.-i
\end{keyword}

\end{frontmatter}

\section{Introduction}
The one-dimensional cubic complex Ginzburg--Landau equation
(CGLE)~\cite{GL} is one of the most-studied nonlinear equations
(see~\cite{review} and references therein). It is a generic
equation which describes many physical phenomena, such as pattern
formation near a supercritical Hopf bifurcation~\cite{review,ch},
the propagation of a signal in an optical
fiber~\cite{AgrawalBook}, spatiotemporal intermittency in
spatially extended dissipative
systems~\cite{vHSvS,MannevilleBook}.

The CGLE
\begin{equation}
\mathrm{i} A_t + p A_{xx} + q |A|^2 A - \mathrm{i} \gamma A =0,
\label{eqCGL3}
\end{equation}
 where subscribes denote partial derivatives: $A_t\equiv\frac{\partial A}{\partial
t}$,  $A_{xx}\equiv\frac{\partial^2 A}{\partial x^2}$, $p\in
\mathbb{C}$, $q\in \mathbb{C}$ and $\gamma \in \mathbb{R}$ is not
integrable if $pq\gamma \neq 0$. In the case $q/p \in \mathbb{R}$,
$\gamma=0$ the CGLE is integrable and coincides with the
well-known nonlinear Schr\"odinger
equation~\cite{Faddeev,Newell,Chernogolovka}.

One of the most important direction in the study of the CGLE is
the consideration of its travelling wave
reduction~\cite{Akhmediev,review,Hecke01,CoMu93,Hecke0110,Hone05,Kudryashov,CoMu03,vSH}:
\begin{equation}
A(x,t)=\sqrt{M(\xi)} \e^{\displaystyle
\mathrm{i}(\varphi(\xi)-\omega t)},\quad \xi=x-ct,\qquad c\in
\mathbb{R}, \quad\omega \in \mathbb{R},\label{travwave}
\end{equation}
which defines the following third order system
\begin{equation}
\left\{
\begin{array}{l}
\displaystyle \frac{M''}{2 M} -\frac{{M'}^2}{4 M^2} - \left(\psi -
\frac{cs_r}{2}\right)^2-
\frac{cs_i M'}{2 M}  + d_r M +g_i=0,\\[2.7mm]
\displaystyle \psi' + \left(\psi- \frac{cs_r}{2}\right)
\left(\frac{M'}{M} - cs_i\right) + d_i M -g_r =0,
\end{array}
\right.
\label{SYSTEM}
\end{equation}
where $\psi\equiv\varphi'\equiv\frac{\d\varphi}{\d\xi}$, 
$M'\equiv\frac{\d M}{\d\xi}$, six real parameters $d_r$, $d_i$, $g_r$,
$g_i$, $s_r$ and $s_i$ are given in terms of $c$, $p$, $q$, $\gamma$ and
$\omega$ as  
\begin{equation}
d_r + \mathrm{i} d_i = \frac{q}{p},\quad s_r - \mathrm{i} s_i =
\frac{1}{p},\quad g_r + \mathrm{i} g_i = \frac{\gamma + \mathrm{i}
\omega}{p} + \frac{1}{2} c^2 s_is_r+\frac{\mathrm{i}}{4}c^2s_r^2.
\end{equation}

Using (\ref{SYSTEM}) one can express $\psi$ in terms of $M$ and
its derivatives:
\begin{equation}
 \psi = \frac{c s_r}{2} + \frac{G'-2 c s_i G}{2
M^2( g_r - d_i M)},
\label{varphi}
\end{equation}
where
\begin{equation}
G\equiv \frac{1}{2} M M'' - \frac{1}{4} M'^2
  -\frac{c s_i}{2} M M' + d_r M^3 + g_i M^2,
\label{GM}
\end{equation}
and obtain the third order equation in $M$:
\begin{equation}
(G'-2 c s_i G)^2 - 4 G M^2 (d_i M - g_r)^2=0. \label{Gequ}
\end{equation}

We will consider the case
\begin{equation}
\frac{p}{q}\not\in \mathbb{R}. \label{pqnotR}
\end{equation}
In this case equation~(\ref{Gequ}) is not
integrable~\cite{CT1989,CoMu93}, which means that the general
solution (which should depend on three arbitrary integration
constants) is not known. Using the Painlev\'e
analysis~\cite{CT1989} or topological arguments~\cite{vSH}  it has been
shown that
single-valued solutions can depend on only one arbitrary
parameter. Equation~(\ref{Gequ}) is autonomous, so this parameter
is $\xi_0$: if $M=f(\xi)$ is a solution, then $M=f(\xi-\xi_0)$,
where $\xi_0\in \mathbb{C}$ has to be a solution. Special
solutions in terms of elementary functions have been found
in~\cite{BN1985,CoMu93,Kudryashov,NB1984}. All known exact
solutions of the CGLE are elementary (rational, trigonometric or
hyperbolic) functions. The full list of these solutions is
presented in~\cite{Hone05,CoMu03}.

In~\cite{CoMu03} a new method to search single-valued particular
solutions has been developed. Rather than looking for an explicit,
closed form expression, R.~Conte and M.~Musette look for the first
order polynomial autonomous ODE for $M(\xi)$. This method allows
to find either elliptic or elementary solutions. It is based on
the Painlev\'e analysis~\cite{Painleve1} and uses the formal
Laurent-series solutions. Using these solutions
A.N.W.~Hone~\cite{Hone05} has proved that a necessary condition
for eq.~(\ref{Gequ}) to admit elliptic solutions is $c=0$. The
goal of this paper is to prove that eq.~(\ref{Gequ}) does not
admit elliptic solutions in the case $c=0$ as well. In other
words, neither travelling nor standing wave solutions are elliptic
functions. In contrast to~\cite{CoMu03,Hone05} we consider
system~(\ref{SYSTEM}) instead of eq.~(\ref{Gequ}). Below we show
that this choice has some preferences. We consider not only
generic (non-zero) values of parameters but also these zero
values. The condition (\ref{pqnotR}) gives only one restriction:
$d_i\neq 0$.

\section{Elliptic functions}

The function $\varrho(z)$ of the complex variable $z$ is a
doubly-periodic function if there exist two numbers $\omega_1$ and
$\omega_2$ with $\omega_1/\omega_2 \not\in \mathbb{R}$, such that
for all $z\in \mathbb{C} $
\begin{equation}
  \varrho(z)=\varrho(z+\omega_1)=\varrho(z+\omega_2).
\end{equation}

By definition a double-periodic meromorphic function is called an
elliptic function. These periods define the period parallelograms
with vertices $z_0$, $z_0+N_1\omega_1$, $z_0+N_2\omega_2$ and
$z_0+N_1\omega_1+N_2\omega_2$, where $N_1$ and $N_2$ are arbitrary
natural numbers and $z_0$ is an arbitrary complex number. The
classical theorems for elliptic functions (see, for
example~\cite{BE,Hurwitz}) prove that

\begin{itemize}

\item If an elliptic function has no poles then it is a constant.

\item The number of elliptic function poles within any finite
period parallelogram  is finite.

\item The sum of residues within any finite period parallelogram
is equal to zero (\textbf{the residue theorem}).

\item If $\varrho(z)$ is an elliptic function then any rational
function of $\varrho(z)$ and its derivatives is an elliptic
function as well.
\end{itemize}

 From (\ref{varphi}) it follows that if $M$ is an
elliptic function then $\psi$ has to be an elliptic function.
Therefore, if we prove that $\psi$ can not be an elliptic
function, we prove that $M$ can not be an elliptic function as
well. To prove this we construct the Laurent-series solutions for
system (\ref{SYSTEM}) and apply the residue theorem to the
function $\psi$ and its degrees.

\section{Nonexistence of the standing wave elliptic solutions}
\subsection{The Laurent-series solutions and the residue theorem}
To prove the non-existence of elliptic solutions to (\ref{SYSTEM})
we will use its solutions in the form of the Laurent series, which
can be easily found due to the Ablowitz--Ramani--Segur algorithm
of the Painleve test~\cite{ARS}. In such a way we obtain solutions
only as a formal series, but really we will use only a finite
number of these series coefficients, so, we do not need the
convergence of these series. It is
known~\cite{CoMu93,Hone05,CoMu03} that there are only two types of
the Laurent-series solutions of (\ref{SYSTEM}) or (\ref{Gequ}).
These solutions depend on only one arbitrary parameter $\xi_0$,
which determines the position of the singular point. At singular
points $\psi$  and $M$ tend to infinity as $1/t$ and $1/t^2$ respectively. We
will use the Laurent series for the function~$\psi$:
\begin{equation}
\psi_1=\sum_{k=-1}^{\infty}C_k(\xi-\xi_0)^k \qquad \mbox{and}
\qquad \psi_2=\sum_{k=-1}^{\infty}D_k(\xi-\tilde{\xi}_0)^k,
\label{psiLaure}
\end{equation}
with $C_{-1}\neq 0$ and  $D_{-1}\neq 0$. A nonconstant elliptic
function should have poles.  Let $\psi(\xi)$ in some parallelogram
of periods has $N_1+N_2$ poles, its Laurent
series expansions are $\psi_1$ in the neighbourhood of $N_1$ poles 
and  are $\psi_2$ in the neighbourhood of $N_2$ poles. 
If $\psi(\xi)$ is an elliptic
function then the sum of its residues in some parallelogram of
periods has to be zero, therefore, this function has both types
of the Laurent series expansions (\ref{psiLaure}) and
\begin{equation}
         N_1=-\,\frac{D_{-1}}{C_{-1}}N_2.
\label{NN}
\end{equation}
If $\psi(\xi)$ is an elliptic function then powers $\psi^k$ have
to be elliptic functions as well, they have $N_1$ Laurent series
expansions $\psi_1^k$ and $N_2$ Laurent series expansions
$\psi_2^k$. To calculate residues of $\psi_1^k$ (or $\psi_2^k$) we
have to use only $k$ leading terms of the Laurent series $\psi_1$
($\psi_2$). The residue theorem for the functions $\psi(\xi)^k$
gives algebraic equations on the coefficients of $\psi_1$ and
$\psi_2$ Laurent series. These series depend on numerical
parameters of system~(\ref{SYSTEM}) and only on them (have no
resonances), hence, we obtain a system of algebraic equations on
coefficients of (\ref{SYSTEM}), at which (\ref{SYSTEM}) can have
elliptic solutions. For example if we demand that the functions
$\psi^2$, $\psi^3$, $\psi^4$ and $\psi^5$ are elliptic, then, using (\ref{NN}),
we obtain the following system on $C_k$ and $D_k$:
\begin{equation}
\left\{
\begin{array}{l}
C_0=D_0,\\
C_1C_{-1}+C_0^2=D_1D_{-1}+D_0^2,\\
C_2C_{-1}^2+3C_1C_0C_{-1}+C_0^3=D_2D_{-1}^2+3D_1D_0D_{-1}+D_0^3,\\
C_3C_{-1}^3+4C_2C_0C_{-1}^2+2C_1^2C_{-1}^2+
6C_{-1}C_0^2C_1+C_0^4={}\\{}=D_3D_{-1}^3+4D_2D_0D_{-1}^2+2D_1^2D_{-1}^2+6D_1D_0^2D_{-1}
+D_0^4.\\
\end{array}
\right.
\label{Lauresys}
\end{equation}
We will use also the corresponding equation for $\psi^7$ under
conditions $C_0=0$, $C_2=0$, $C_4=0$, $D_0=0$, $D_2=0$ and $D_4=0$:
\begin{equation}
C_5C_{-1}^5+6C_3C_1C_{-1}^4+5C_1^3C_{-1}^3=D_5D_{-1}^5+6D_{-1}^4D_3D_1+
5D_{1}^3D_{-1}^3.
\label{psi7}
\end{equation}
We have calculated the residues of powers of $\psi$ with the help
of the procedure \textsf{ydegree} from our package of
Maple~\cite{Maple} procedures~\cite{VernovCASC}, which realizes
the Conte--Musette algorithm for construction of single-valued
solutions of nonintegrable systems~\cite{CoMu03}.

\subsection{The number of essential numerical parameters of system~(\ref{SYSTEM})}
System~(\ref{SYSTEM}) includes seven arbitrary constants, some of
them can be fixed without loss of generality. First of all one can
fix $s_r$ and $s_i$. From the condition $p \not\in \mathbb{R}$
(the case of real $p$ we consider separately) follows that
$s_i\neq 0$. Using the following transformations:
\begin{equation}
\tilde{c}=\varpi c, \qquad  \tilde{s}_i=\frac{s_i}{\varpi}, \qquad
\tilde{s}_r=\tau s_r, \qquad
\tilde{\psi}=\psi-\frac{cs_r}{2}(1-\tau\varpi)
\end{equation}
one can put
\begin{equation}
 \tilde{s}_r={}-\frac{1}{10} \qquad\mbox{and}\qquad
\tilde{s}_i={}-\frac{3}{10}. \label{srsifix}
\end{equation}

Using the transformations
\begin{equation}
\tilde{M}=\mu M, \qquad  \tilde{d}_i=\frac{d_i}{\mu}, \qquad
\tilde{d}_r=\frac{d_r}{\mu}
\end{equation}
we can fix $d_r$ or $d_i$. Following~\cite{Hone05} we will fix the
value of $d_r$. Our restriction (\ref{pqnotR}) on parameters $p$ and 
$q$ gives no information about $d_r$, so we have to consider two cases: $d_r=0$
and $d_r\neq 0$ separately. Using scaling transformations of the
independent variable $\xi$ it is possible to fix $g_i$ or $g_r$,
but, following~\cite{CoMu03,Hone05} we leave them arbitrary to
consider zero and nonzero values of these parameters at once.

From the second equation of (\ref{SYSTEM}) it follows that if
$\psi$ is a constant then $M$ can not be an elliptic function, so
to obtain nontrivial elliptic solutions we have to assume that
$\psi$ has poles. We do not restrict ourself to the case $c=0$ and
prove the non-existence of either travelling or standing wave
solutions. It has been noted in~\cite{CoMu04} that one does not
need to transform a system of differential equations into one
equation to obtain the Laurent-series solutions.

\subsection{The case $d_r=0$}
Let us consider system~(\ref{SYSTEM}) with
\begin{equation}
d_r=0, \qquad s_r={}-\frac{1}{10} \qquad\mbox{and}\qquad
s_i={}-\frac{3}{10}. \label{dr0}
\end{equation}

From the condition (\ref{pqnotR}) it follows
that $d_i\neq 0$, therefore, there exist two different
Laurent-series solutions ($\xi_0=0$) of~(\ref{SYSTEM}):
\begin{equation}
\breve{\psi}_1=\frac{\sqrt {2}}{\xi}-\frac {c(\sqrt{2}+1)}{20}+
\mathcal{O}(\xi), \label{psi1}
\end{equation}
\begin{equation}
\breve{M}_1=\frac{3\sqrt {2}}{d_i}\left(\frac{1}{\xi^2}-\frac
{1}{10\xi}\right) + \mathcal{O}(1) \label{M1}
\end{equation}
and
\begin{equation}
\breve{\psi}_2=-\frac{\sqrt{2}}{\xi}+\frac {c(\sqrt{2}-1)}{20}+
\mathcal{O}(\xi), \label{psi2}
\end{equation}
\begin{equation}
\breve{M}_2=-\frac{3\sqrt{2}}{d_i}\left(\frac{1}{\xi^2}-\frac
{1}{10\xi}\right) + \mathcal{O}(1) \label{M2}
\end{equation}

From (\ref{NN}) it follows that $N_1=N_2$, that is to say, if the
function $\psi(\xi)$ has $N$ poles with residues, which are equal
to $\sqrt{2}$, within some finite period parallelogram, then in
this domain the number of poles, which residues are equal to
$-\sqrt{2}$, has to be equal to $N$ as well.

Residues of $\breve{\psi}_1^2$ are equal to
$-2\sqrt{2}c(\sqrt{2}+1)/20$, whereas residues of
$\breve{\psi}_2^2$ are $-2\sqrt{2}c(\sqrt{2}-1)/20$. From the
first equation of system (\ref{Lauresys}) we obtain that the sum
of residues of the function $\psi^2$ is equal to zero if and only
if $c=0$. So, we prove the absence of the travelling wave
solutions. Note that to obtain this result we have used only two
coefficients of the Laurent series $\psi_1$ and $\psi_2$.  In the
case $c=0$ we have to apply the residue theorem for $\psi_3$ and
$\psi_4$, so, we have to calculate four coefficients in these
series (two of them are zero at $c=0$)
\begin{equation}
\breve{\psi}_1=\frac{\sqrt {2}}{\xi}+\frac{0}{\xi}+\frac
{1}{21}\,\left(5\sqrt{2}g_i-g_r\right)\xi +0\xi^2+
\mathcal{O}(\xi^3), \label{psi1f}
\end{equation}
and
\begin{equation}
\breve{\psi}_2=-\frac{\sqrt{2}}{\xi}+\frac{0}{\xi} -{\frac
{1}{21}}\,\left(5\sqrt {2}g_i+g_r\right)\xi +0\xi^2+
\mathcal{O}(\xi^3), \label{psi2f}
\end{equation}

From the second and the third equations of (\ref{Lauresys}) we
obtain that the functions $\psi^3$ and $\psi^4$ satisfy the
residue theorem if and only if
\begin{equation}
g_i=0 \qquad \mbox{and} \qquad g_r=0.
\label{gg00}
\end{equation}

In this case the Laurent-series solutions give
\begin{equation}
\breve{\psi}_1(\xi)=\frac{\sqrt{2}}{\xi},  \qquad
\breve{M}_1(\xi)=\frac{3\sqrt{2}}{d_i\xi^2}
\end{equation}
and
\begin{equation}
\breve{\psi}_2(\xi)={}-\frac{\sqrt{2}}{\xi}, \qquad  
  \breve{M}_2(\xi)={}-\frac{3\sqrt{2}}{d_i\xi^2}.
\end{equation}

The straightforward substitution of these functions in
system~(\ref{SYSTEM}) with $c=0$, $d_r=0$, $g_r=0$ and $g_i=0$
proves that they are exact solutions. Using the
Ablowitz--Ramani--Segur algorithm~\cite{ARS} it is easy to prove
that the coefficients of the Laurent-series solutions does not
include arbitrary parameters, so the obtained solutions are unique
single-valued solutions and the CGLE has no elliptic solution for
these values of parameters as well. Thus we have proved the
non-existence of both travelling and standing wave elliptic
solutions at $d_r=0$.

\subsection{The case $d_r\neq 0$}

In this case we can use the following values of parameters without
loss of generality
\begin{equation}
d_r=\frac{1}{2}, \qquad s_r={}-\frac{1}{10} \qquad\mbox{and}\qquad
s_i={}-\frac{3}{10}. \label{dr1_2}
\end{equation}

To simplify calculations we, following~\cite{Hone05}, express
$d_i$ through a new real parameter:
\begin{equation}
d_i=\pm \frac{3\sqrt{\beta^2-1}}{4\sqrt{2}}, \qquad \beta>1.
 \label{dibeta}
\end{equation}

If $d_i>0$ (the sign $+$ in (\ref{dibeta})) then
system~(\ref{SYSTEM}) has the following Laurent series solutions:
\begin{equation}
\tilde\psi_1=\frac{\sqrt{2}(\beta + 1)}{\sqrt{\beta^2 -
1}}\xi^{-1}-\frac{c}{10}\left(\frac{\sqrt{\beta^2 -
1}}{\sqrt{2}(\beta - 1)}+\frac{1}{2}\right)+\mathcal{O}(\xi)
\label{psi3}
\end{equation}
\begin{equation}
\tilde\psi_2=-\frac{\sqrt{2}(\beta - 1)}{\sqrt{\beta^2 -
1}}\xi^{-1}+\frac{c}{10}\left(\frac{\sqrt{\beta^2 -
1}}{\sqrt{2}(\beta +1)}-\frac{1}{2}\right)+ \mathcal{O}(\xi)
\label{psi4}
\end{equation}

If there are $N_1$ Laurent series of type $\tilde\psi_1$ and $N_2$
Laurent series of type $\tilde\psi_2$ then  eq.~(\ref{NN}) gives
\begin{equation}
N_1=\frac{\beta-1}{\beta+1}N_2. \label{N1N2beta}
\end{equation}
The residues theorem for $\psi^2$ gives $c\beta=0$. Using condition
$\beta>1$, we derive that $c$ has to be equal to zero, 
so we reobtain the main result of~\cite{Hone05} 
that the CGLE has no elliptic travelling wave solutions for non-zero values of
parameters. Note that the use of Laurent series of $\psi(\xi)$
instead of the Laurent series of $M(\xi)$ allows to simplify calculations.

Let us consider the standing wave solutions ($c=0$) of the CGLE.
The Laurent series solutions are:
\begin{equation}
\begin{array}{l}
\displaystyle\tilde{\psi}_1={\frac {\sqrt {2} \left(\beta+1 \right)
}{\sqrt{\beta^2-1}}}\,\xi^{-1}-{\frac{\left( \beta
g_r-5g_r-5\sqrt{2(\beta^2-1)}g_i \right)
}{3(7\,\beta+5)}}\,\xi+{}\\[2.7mm]
\displaystyle{}+\frac{1}{90(\beta+1)(7\beta+5)^2}\left\{32\beta^3
g_i g_r-256\beta^2g_ig_r-32\beta
g_i g_r+{}\right.\\[2.7mm]
\displaystyle {}+256g_i g_r +\sqrt{2(\beta^2-1)}\left(
122\beta^2g_i^2+11\beta^2g_r^2
-34\beta g_r^2+{}\right.\\[2.7mm]
\displaystyle\left.\left.{}+61g_r^2-122g_i^2\right) \right\}\,
\xi^3+\mathcal{O}(\xi^5),
\end{array}
 \label{psi1c0}
\end{equation}

\begin{equation}
\begin{array}{l}
\displaystyle\tilde{\psi}_2=-{\frac{\sqrt{2}\left(\beta
-1\right)}{\sqrt {{\beta}^{2}-1}}}\xi^{-1}- {\frac { \left(
\beta g_r+5g_r+5\sqrt{2(\beta^2-1)}g_i \right)}{3(7\,\beta-5)}}\xi+{}\\[2.7mm]
\displaystyle{}+\,\frac{1}{90(\beta+1)(7\beta+5)^2}\left\{
 32\beta^3g_i\,g_r+256\,{\beta}^{2}g_i\,g_r-32\beta g_i g_r-{}\right. \\[2.7mm]
{}-256g_ig_r+
 \sqrt{2(\beta^2-1)}\left(122\beta^2g_i^2-11\beta^2g_r^2
+34\beta g_r^2+{}\right.\\[2.7mm]
\displaystyle\left.\left.{}+61g_r^2-122g_i^2\right) \right\}\,
\xi^3+\mathcal{O}(\xi^5).
\end{array}
 \label{psi2c0}
\end{equation}

The residues of $\psi^2$, $\psi^4$  and $\psi^6$ are equal to zero
at $c=0$. Substituting the coefficients of the Laurent series
$\tilde\psi_1$ and $\tilde\psi_2$,  we transform
system~(\ref{Lauresys}) and eq.~(\ref{psi7}) into the algebraic
system in $\beta$, $g_i$ and $g_r$. This system is too cumbersome
to be presented here. The condition $\beta>1$ leaves only one
solution of this system:
\begin{equation}
g_r=0, \qquad g_i=0. \label{grgi0}
\end{equation}

In the case $d_i<0$ we also obtain that the residue theorem for
powers of $\psi$ can be satisfied only if $g_r=0$ and $g_i=0$. Let
us consider system~(\ref{SYSTEM}) with zero values of $c$, $g_i$
and $g_r$, $d_r=1/2$ and an arbitrary (nonzero) value of $d_i$:
\begin{equation}
\left\{
\begin{array}{l}
\displaystyle 2MM'' -{M'}^2 - 4M^2\psi^2+ 2M^3=0,\\[2.7mm]
\displaystyle M\psi' + M'\psi + d_i M^2 =0,
\end{array}
\right. \label{SYSTEM0}
\end{equation}

The straightforward substitution gives that functions
\begin{equation}
\tilde{\psi}_1(\xi)=\frac{3+\sqrt{9+32d_i^2}}{4d_i\xi}, \qquad \tilde{M}_1(\xi)
=\frac{3(3+\sqrt {9+32d_i^2})}{4d_i^2\xi^2}
\end{equation}
and
\begin{equation}
\tilde{\psi}_2(\xi)=\frac{3-\sqrt{9+32d_i^2}}{4d_i\xi}, 
\qquad \tilde{M}_2(\xi)=\frac
{3(3-\sqrt {9+32d_i^2})}{4d_i^2\xi^2}
\end{equation}
are exact solutions of system~(\ref{SYSTEM0}). This system has no
other single-valued solutions, so we have proved the non-existence
of neither elliptic standing wave nor elliptic travelling wave
solutions in case $d_r\neq 0$ as well. In our calculations we
assume that $s_i\neq 0$. At the same time our results prove the
non-existence of elliptic solutions in the case $s_i=0$ too.
Indeed, if $c=0$ then cases with $s_i=0$ and $s_i\neq 0$
coincide, to transform the case \{$s_i=0$, $c\neq 0$\} into the
considered case \{$s_i\neq 0$, $c=0$\} we have to add a constant
to $\psi(\xi)$.

\section{Conclusions}

The Laurent-series solutions are useful not only to find elliptic
solutions, but also to prove the non-existence of them. Using the
Hone's method, based on residue theorem, we have proved the
non-existence of both standing and travelling wave elliptic
solutions of the CGLE in the case $p/q\not\in\mathbb{R}$. Our
result amplifies the Hone's result~\cite{Hone05}, that the CGLE
with generic (non-zero) values of parameters has no elliptic
travelling wave solution.

\section*{Acknowledgements}
The author is grateful to  \  R.~Conte, \ who attracted his
attention to the paper~\cite{Hone05}, and  \  A.N.W. Hone  \  for useful
comments. This work has been supported in part by Russian
Fede\-ration President's Grant NSh--1685.2003.2 and by the grant
of  the scientific Program "Universities of Russia" UR.02.02.503.

%=============================================================


\begin{thebibliography}{72}

\bibitem{ARS} M.~J.~Ablowitz, A.  Ramani, H.  Segur,
A Connection between Nonlinear Evolution Equations and Ordinary
Differential Equations of P-type. I $\&$ II, {\em  J. Math.
Phys.\/} {\bf 21} (1980) 715--721, $\&$ 1006--1015.

\bibitem{AgrawalBook} G.~P.~Agrawal,
{\em Nonlinear fiber optics,\/} Academic press, Boston, 1989.

\bibitem{Akhmediev} N. N.~Akhmediev, V. V.~Afanasjev,
J.~M.~Soto-Crespo, Singularities and spesial soliton solutions of
the cubic-quintic complex Ginzburg--Landau equation, {\em Rev.
Phys. E\/} {\bf 53} (1996) 1190--1201.
\bibitem{review} I. Aranson, L. Kramer, The World of the
Complex Ginzburg--Landau Equation, {\em Rev. Mod. Phys.\/} {\bf
74} (2002) 99--143, [cond-mat/0106115].

\bibitem{BN1985} N.~Bekki, K.~Nozaki,
Formations of spatial patterns and holes in the generalized
Ginzburg--Landau equation, {\em  Phys.~Lett.~A\/} {\bf 110} (1985)
133--135.

\bibitem{Hecke01}  L.~Brusch, A.~Torcini, M.~van~Hecke, M.~G.~Zimmermann, M.~B\"ar,
Modulated Amplitude Waves and Defect Formation in the
One-Dimensional Complex Ginzburg--Landau Equation, {\em Physica
D\/} {\bf  160} (2001) 127--148, [nlin.CD/0104029].

\bibitem{CT1989} F.~Cariello, M.~Tabor,
Painlev\'e expansions for nonintegrable evolution equations, {\em
Phisica D\/} {\bf 39} (1989) 77--94.

\bibitem{CoMu93} R.~Conte, M.~Musette,
Linearity inside nonlinearity: exact solutions to the complex
Ginzburg-Landau equation, {\em Phisica D\/} {\bf 69} (1993) 1--17.


\bibitem{CoMu04} R. Conte, M. Musette, Solitary waves of nonlinear
nonintegrable equations, nlin.PS/0407026.

\bibitem{ch} M. C. Cross, P. C. Hohenberg, Pattern formation outside
of equilibrium, {\em  Rev. Mod. Phys.\/} {\bf 65} (1993) 851--1112.

\bibitem{GL} V. L. Ginzburg, L. D. Landau, On the theory of superconductors,
{\em Zh. Eksp. Teor. Fiz. (Sov. Phys. JETP)\/} {\bf 20} (1950)
1064--1082; English translation in L.~D.~Landau, {\em Collected
Papers\/}, Oxford, Pergamon Press, 1965, p.~546.

\bibitem{BE}
 A.  Erd\'elyi et al. (eds.), {\em Higher
 Transcendental
Functions (based, in part, on notes left by H.~Bateman),  Vol. 3
\/}, MC Graw-Hill Book Company, New York, Toronto, London, 1955.

\bibitem{Faddeev} L. D. Faddeev, L. A. Takhtajan, {\em Hamiltonian Methods in
the Theory of Solitons\/}, "Nauka", Moscow, 1986; English
translation: Springer-Verlag \{Springer series in Soviet
mathematics\}, Berlin, 1987.

\bibitem{Maple} A.~Heck, {\em Introduction
to Maple, 3rd Edition\/}, Springer--Verlag, New York, 2003.

\bibitem{vHSvS} M.~van Hecke, C.~Storm, W.~van Saarlos,
Sources, sinks and wavenumber selection in coupled CGL equations
and experimental implications for counter-propagating wave
systems, {\em Phisica D\/}  {\bf 133} (1999) 1--47,
[Patt-sol/9902005].

\bibitem{Hecke0110} M.~van Hecke, Coherent and Incoherent structures in systems
described by the 1D CGLE: Experiments and Identification, {\em
Phisica D\/} {\bf 174} (2003) 134--151, [cond-mat/01100068].

\bibitem{Hurwitz}
 von A. Hurwitz, {\em Allgemeine Funktionentheorie
und Elliptische Funktionen\/}, von R. Courant, {\em Geometrische
Funktionentheorie\/}, Springer--Verlag, Berlin, New York, 1964.

\bibitem{Hone05} A. N. W. Hone, Non-existence of elliptic
travelling wave solutions of the complex
Ginzburg--Landau equation, {\em Phisica D\/} (2005) in press.

\bibitem{Kudryashov} N.~A.~Kudryashov,
Exact solutions of a generalized equation of Ginzburg--Landau,
{\em Matematicheskoye modelirovanie\/} {\bf 1} (1989) 151--158
\{in Russian\}.

\bibitem{MannevilleBook} P.~Manneville,
{\em Dissipative structures and weak turbulence} (Academic Press,
Boston, 1990). French adaptation: \textit{Structures dissipatives,
chaos et turbulence}, Al\'ea-Saclay, Gif-sur-Yvette, 1991.

\bibitem{CoMu03} M. Musette, R. Conte,
Analytic solitary waves of nonintegrable equations, {\em Phisica
D\/} {\bf 181} (2003) 70--79, [nlin.PS/0302051].

\bibitem{Newell} A.~C.~Newell, {\em Solitons in Mathematics and
Physics\/}, Society for Industrial and Applied Mathematics,
Filadelphia, 1985.

\bibitem{NB1984} K.~Nozaki, N.~Bekki,
Exact solutions of the generalized Ginzburg--Landau equation, {\em
J.~Phys.~Soc.~Japan\/}  {\bf 53} (1984) 1581--1582.

\bibitem{Painleve1}
P.~Painlev\'e,  {\em  Le\c{c}ons sur la th\'eorie analytique des
\'equations diff\'e\-rentielles, profees\'ees \`a Stockholm
(septembre, octobre, novembre 1895) sur l'invitation de S. M. le
roi de Su\`ede et de Norw\`ege\/},  Hermann, Paris, 1897;
Reprinted in:  {\em  O$\!$euvres\ de Paul Painlev\'e, V.~1\/}, ed.
du CNRS, Paris, 1973. On-line version: The Cornell Library
Historical Mathematics Monographs,
\verb$http://historical.library.cornell.edu/$

\bibitem{Chernogolovka} The Proceedings of the Conference
on the Nonlinear Schr\"odenger Equation, Chernogolovka, Russia,
1994, {\em  Phisica D\/} {\bf 87} (1995) 1--380.

\bibitem{vSH} W.~van Saarloos, P.~C.~Hohenberg,
Fronts, pulses, sources and sinks in generalized complex
Ginzburg--Landau equations, {\em Phisica D\/} {\bf 56} (1992)
303--367, Erratum {\bf 69} (1993) 209.

\bibitem{VernovCASC} S.~Yu. Vernov, Construction of
Single-valued Solutions for Nonintegrable Systems with the Help of
the Painlev\'e Test, in: V.~G.~Ganzha, E.~W.~Mayr,
E.~V.~Vorozhtsov (Eds.), {\em Proceedings of the International
Conference "Computer Algebra in Scientific Computing"\/} (St.
Petersburg, Russia, 2004), Technische Universitat, Munchen,
Garching, 2004, pp.~457--465, [nlin.SI/0407062].

\end{thebibliography}
\end{document}